\documentclass[12pt]{article}

\pdfoutput=1

\usepackage{amsmath,amssymb,amsfonts}
\usepackage{color}
\definecolor{darkblue}{rgb}{0.1,0.1,.7}
\usepackage[colorlinks, linkcolor=darkblue, citecolor=darkblue, urlcolor=darkblue, linktocpage]{hyperref} 
\usepackage[square, comma, compress,numbers]{natbib}
\usepackage[]{version}
\usepackage[]{amsmath}
\usepackage[]{graphicx}
\usepackage[]{latexsym}
\usepackage{geometry}
\usepackage{amscd}
\usepackage[all,cmtip]{xy}
\usepackage{mathrsfs}
\usepackage[margin=10pt,font=small,labelfont=bf]{caption}
\numberwithin{equation}{section}

\newcommand{\reef}[1]{(\ref{#1})}

\newcommand{\be}{\begin{equation}}
\newcommand{\ee}{\end{equation}}
\newcommand{\bea}{\begin{eqnarray}}
\newcommand{\eea}{\end{eqnarray}}

\newcommand{\vareps}{\varepsilon}
\newcommand{\eps}{\epsilon}
\def\beq{\begin{equation}} 
\def\eeq{\end{equation}} 
\def\del {\partial} 
\def\nn{\nonumber}

\def\calO {{\cal O}}

\def\ge{\geqslant}

\begin{document}

\vspace*{-.6in} \thispagestyle{empty}
\begin{flushright}
LPTENS--12/35\\
CERN-PH-TH/2012-324
\end{flushright}
\vspace{.2in} {\Large
\begin{center}
{\bf The $S$ parameter for a Light Composite Higgs:\\ a Dispersion Relation Approach}
\end{center}
}
\vspace{.2in}
\begin{center}
{\bf 
Axel Orgogozo$^{a}$ and Slava Rychkov$^{b,a,c}$
}
\\
\vspace{.6in}  
{$^a$ Laboratoire de Physique Th\'{e}orique, \'{E}cole Normale Sup\'{e}rieure, Paris, France\\
$^b$ CERN, Theory Division, CH-1211 Geneva, Switzerland\\ 
$^c$ Facult\'{e} de Physique, Universit\'{e} Pierre et Marie Curie, Paris, France}
\end{center}

\vspace{.2in}

\begin{abstract}
We derive a dispersion relation for the $S$ parameter in the $SO(5)/SO(4)$ Minimal Composite Higgs model. This generalizes the Peskin-Takeuchi formula to the case when a light Higgs boson is present in the spectrum. Our result combines an IR effect due to the reduction in the Higgs boson couplings with a UV contribution from the strong sector. It also includes a finite matching term, achieving a very good relative accuracy $O(m_h/m_\rho)$. We apply our formula in several toy examples, modeling the UV spectral density via Vector Meson Dominance.  
\end{abstract}
\vspace{.2in}
\vspace{.3in}
%\vskip 1cm 
\hspace{0.7cm} November 2012

\newpage

\tableofcontents

\section{Introduction}

The only two known mechanisms to protect the Higgs boson mass from large quantum corrections are supersymmetry and shift symmetry. Here we focus on the second possibility, imagining that the newly discovered Higgs boson is a pseudo-goldstone boson of a spontaneously broken global symmetry \cite{Kaplan:1983fs,Kaplan:1983sm}. This class of models are known as `composite Higgs'. Our purpose is to discuss, in this context, the leading $O(g^2)$ oblique contribution to the $S$ parameter. Computations of electroweak precision observables in the Standard Model (SM) and its extensions is of course an old and well-studied problem. This problem is nontrivial for composite Higgs models due to strong interactions present in the EWSB sector. For this reason the electroweak parameters cannot be computed in a purely perturbative approach. Rather, a combination of perturbation theory with dispersive techniques is needed. In fact, our main result, Eq.~\reef{eq:main}, can be seen as a generalization of the Peskin-Takeuchi dispersion relation for the $S$ parameter \cite{Peskin:1991sw} from the Higgsless to the composite Higgs case.

For concreteness, we will consider here only the Minimal Composite Higgs Model (MCHM) \cite{Agashe:2004rs}. This model assumes the $SO(5)/SO(4)$ pattern of global symmetry breaking. Other, more goldstone-rich, patterns can be considered analogously. 
The effective low energy description of the MCHM is via a real scalar five-plet $\Phi$ subject to a constraint $\Phi^2=f^2$. The low energy spectrum contains, in addition to three goldstones eventually eaten by the $W$ and $Z$, a fourth (pseudo)goldstone which is the composite Higgs boson. The $SO(4)=SU(2)_L\times SU(2)_R$ subgroup of the global symmetry acting on the first four components of $\Phi$ plays the role of the custodial group; its $SU(2)_L\times U(1)_{T^3_R}$ subgroup is gauged by the SM gauge group. The strength of the electroweak symmetry breaking is then controlled by the vacuum orientation: 
\beq
\langle\Phi\rangle=(0,0,0,f \sin\theta ,f\cos\theta).
\label{eq:mis}
\eeq
A potential for $\theta$ will be generated by small $SO(5)$ breaking effects, such as couplings to the SM gauge fields and fermions. The low energy dynamics of the model is thus captured by the Lagrangian
\beq
\frac 12(D_\mu \Phi)^2+V(\theta)\,.
\label{eq:leeff}
\eeq
To correctly reproduce the electroweak scale, the minimum of the potential must be
at
\beq
f \sin\theta_*=v=246 \text{ GeV}.
\eeq
The Higgs boson field corresponds to fluctuations in $\theta$ around the minimum and its mass is given by
\beq
m_h^2=f^2 V''(\theta_*)\,.
\eeq
Only these two characteristics of the potential will be important below.

In this class of models, the $S$ parameter gets contributions from two effects. The first one (which is not specific to the $SO(5)/SO(4)$ pattern but is present for any composite Higgs model) is associated with the strong sector spin one resonances of mass $m_\rho \sim g_*f$, where $g_*\lesssim 4\pi$ is the strong sector interaction strength. This contribution, called UV, can be estimated as\footnote{The
parameters $\hat{S},\hat{T}$ \cite{Barbieri:2004qk} are proportional to the
Peskin-Takeuchi parameters: $\hat{S}=\frac
{g^{2}}{16\pi}S$, $\hat{T}=\alpha_{\mathrm{EM}}T$.}
\beq
\hat S_{\rm UV}\sim \frac{m_W^2}{m_\rho^2}\,.
\label{eq:SUV}
\eeq
In 5D models realizing composite Higgs dynamics (or deconstructions thereof) this contribution can be computed integrating out the resonances at tree level.

The second effect on the $S$ parameter is associated with the fact that the composite Higgs boson couplings to $WW$ and $ZZ$ are suppressed with respect to the SM by a factor of $a\equiv\cos\theta_*<1$ \cite{Giudice:2007fh}. Since the Higgs decay signal into $WW$ and $ZZ$ observed by the LHC experiments is roughly consistent with the SM rate, this suppression cannot be too large. A conservative lower bound is $a\gtrsim0.8$ at $90\%$ CL.\footnote{This bound follows combining the latest reported $WW$ and $ZZ$ signal strengths \cite{CMS-PAS-HIG-12-045,ATLAS-CONF-2012-162}. We assumed for simplicity that the Higgs boson couplings to the fermions are as in the SM; see \cite{Azatov:2012ga,Giardino:2012dp,Ellis:2012hz,Montull:2012ik,Espinosa:2012im} for more detailed fits.}, implying $f\gtrsim400$ GeV.

As pointed out in \cite{Barbieri:2007bh}, $a<1$ induces a change in the electroweak oblique parameters. The change is easy to write down for the heavy Higgs $m_h\gg m_Z$. In this approximation, the Higgs mass dependence of the $S$ and $T$ parameters in the Standard Model is given by 
\beq
\hat S, \hat T\approx \kappa_{S,T}\log\frac {m_h}{m_Z}\,,\qquad \kappa_S=\frac{g^2}{96\pi^2}\,,\quad \kappa_T=-\frac{3g'^2}{32\pi^2}\,.
\label{eq:STheavy}
\eeq 
The origin of these finite logarithms is a cancellation of the Higgs boson loop against the goldstone boson loop, which are separately logarithmically divergent:
\beq
\log\frac {m_h}{m_Z}=\left(\log\frac{\Lambda}{m_Z}\right)_{\rm goldstones}-\left(\log\frac{\Lambda}{m_h}\right)_{\rm Higgs}\,,
\eeq
where $\Lambda$ is a UV cutoff.
For composite Higgs, the second logarithm is rescaled by $a^2$, so that the logarithmic divergence is not canceled completely.

The oblique parameters can then be computed by \reef{eq:STheavy} provided that the Higgs mass is replaced by an effective mass
\beq
m_{\rm eff}=m_h\left(\frac\Lambda{m_h}\right)^{1-a^2}\,.
\eeq
The full $S$ parameter is obtained by adding the UV contribution \reef{eq:SUV} and the IR contribution from \reef{eq:STheavy}
with $m_h\to m_{\rm eff}$ and $\Lambda\sim m_\rho$. This prescription was first advocated in \cite{Barbieri:2007bh} and used in a number of subsequent works \cite{Gillioz:2008hs,Anastasiou:2009rv,Galloway:2010bp,Espinosa:2010vn,Panico:2010is,Marzocca:2012zn,Bellazzini:2012tv,Barbieri:2012tu}. 

A new computation of the $S$ parameter proposed below is better than the above prescription in several aspects:
\begin{itemize}
\item
The UV contribution is computed via a dispersion integral rather than estimated via \reef{eq:SUV}. 
\item 
The IR contribution is computed without using the heavy Higgs approximation, as required in view of the rather low experimentally observed $m_h\approx 125$ GeV.
\item
 A finite matching term is computed, allowing to put together the UV and IR contributions and eliminate cutoff ambiguities.
\end{itemize}
Our final formula determines the $S$ parameter with relative accuracy $O(m_h/m_\rho)$. 

The paper is organized as follows. In section \ref{sec:SM} we remind how oblique parameters depend on the Higgs mass in the SM, beyond the heavy Higgs approximation. In section \ref{sec:MCHM} we discuss the $S$ parameter in the minimal composite Higgs model. Our main result here is a formula for the complete $S$ parameter at $O(g^2)$ beyond the heavy Higgs limit. Our formula combines, in a cutoff-independent way, the IR composite Higgs contribution with a dispersive integral computing the UV contribution of resonances. This formula plays a role similar to the Peskin-Takeuchi dispersion relation in Technicolor models \cite{Peskin:1991sw}. We then discuss in section \ref{sec:toy} several toy model applications of our formula, using Vector Meson Dominance to parametrize the unknown spectral densities, and conclude. Our group theory conventions are gathered in the appendix.

%In section ?? we discuss the $T$ parameter. We give the IR composite Higgs contribution, but at present we don't know how to combine it with the UV contribution computed in terms of the strong theory data. 

\section{Oblique parameters in the Standard Model}
\label{sec:SM}
In this section we review the Higgs boson mass dependence of the electroweak precision observables in the SM.
A useful basis is given by the three $\vareps$ parameters \cite{Altarelli:1990zd,Altarelli:1991fk}, particular linear combinations of the precision observables $\Delta \rho$, $\Delta k$, and $\Delta r_\text{w}$. The $\vareps$'s can be computed as follows:
\begin{equation} 
\begin{split}
\varepsilon_1 &= e_1-e_5+\text{non-oblique}\,,\\
\varepsilon_2 &= e_2-c^2e_5-(s^2e_4+\text{non-oblique})\,,\\
\varepsilon_3 &= e_3-c^2e_5+(c^2e_4+\text{non-oblique})\,,
\label{eq:epsi}
\end{split}
\end{equation}
where the $e$'s are expressed in terms of the electroweak gauge boson self-energies:\begin{equation} 
\begin{split}
e_1 & =\frac{A_{33}(0)-A_{WW}(0)}{m_W^2} =\frac{A_{ZZ}(0)}{m_Z^2}-\frac{A_{WW}(0)}{m_W^2}\,,\\
e_2 & =F_{WW}(m_W^2)-F_{33}(m_Z^2)\,,\\
e_3 & =\frac{c}{s}F_{30}(m_Z^2)\,,\\
e_4 & =F_{\gamma\gamma}(0)-F_{\gamma\gamma}(m_Z^2)\,,\\
e_5 & =m_Z^2 F_{ZZ}'(m_Z^2)\,.
\end{split}
\label{es}
\end{equation}
Here $F(q^2)$ and $A(0)$ appear in the decomposition of the polarization operators: 
\beq
\Pi_{ij}^{\mu\nu}(q)\equiv-i\eta^{\mu\nu}\left[A_{ij}(0)+q^2F_{ij}(q^2)\right]+q^{\mu}q^{\nu}\ \text{terms}\,.
\eeq
The $s$ and $c$ in \reef{eq:epsi} are the sine and cosine of the Weinberg angle. Finally, the `non-oblique' terms involve the vertex and box corrections. These terms, as well as $e_4$ with which they are grouped, do not involve the Higgs boson and we won't need to discuss them.

We have evaluated the one loop diagrams contributions to $e_{1,2,3,5}$ involving the Higgs boson. The relevant diagrams involve the Higgs and a goldstone, or the Higgs and a gauge boson propagating in the loop. The Higgs dependent parts of the $\vareps$ parameters are then found to be:
\begin{align}
\varepsilon_{1,\text{Higgs}}&\equiv(e_1-e_5)_\text{Higgs}=\frac{3g^2}{64\pi^2c^2} H_A(h)+\frac{3g'^2}{64\pi^2}\log\frac{\Lambda^2}{m_Z^2}+\frac{g^2}{384\pi^2 c^2}\left(-4+3c^2+18c^2\log c^2\right)\,,\nn\\
\varepsilon_{2,\text{Higgs}}&\equiv(e_2-c^2e_5)_\text{Higgs}\nn\\
&\quad=\frac{3g^2}{64\pi^2(c^2-s^2)}\left\{[H_A(h)-H_m(h)]-2s^2[H_A(h)-H_R(h)]\right\}
-\frac{g^2}{384\pi^2}(1+2\log{c^2})\,,\nn\\
\varepsilon_{3,\text{Higgs}}&\equiv(e_3-c^2e_5)_\text{Higgs}=\frac{3g^2}{64\pi^2}\left[H_A(h)-H_R(h)\right]-\frac{g^2}{192\pi^2}\log\frac{\Lambda^2}{m_Z^2}-\frac{g^2}{288\pi^2}\,.
\label{eq:epsHiggs}
\end{align}
We computed this result in dimensional regularization with
\beq
\log \Lambda^2\equiv N_\eps+\log \mu^2\,,\quad N_\eps\equiv  \frac{2}{\epsilon}+\log (4\pi)-\gamma+1\,.
\label{cutoff}
\eeq
We also did it in an arbitrary $\xi$-gauge. The result \reef{eq:epsHiggs} is $\xi$-independent, for the following reason. The total $\vareps$'s should of course be gauge-independent. Moreover, the $\xi$-dependent terms in the diagrams involving the Higgs and in the diagrams involving only the gauge bosons cannot cancel each other since the former depend on the Higgs mass and the latter do not. So these two groups of $\xi$-dependent terms should cancel separately.\footnote{We thank Adam Falkowski for discussions concerning this point.}

To facilitate comparison with the literature, we expressed \reef{eq:epsHiggs} using the functions $H_i(h)$ of $h= m_h^2/m_Z^2$ given by
\begin{align}
H_A(h)=&\frac{h c^2}{h-c^2}\log \frac{h}{c^2}-\frac{8h}{9(h-1)}\log h+ \left(\frac{4}{3}-\frac{2}{3}h+\frac{2}{9}h^2\right)F_h(h)-\left(\frac{4}{3}-\frac{4}{9}h+\frac{h^2}{9}\right)F_h'(h)-\frac{h}{18}\,,\nn\\
H_R(h)=&-\frac{h}{18}+\frac{c^2}{1-\frac{c^2}{h}}\log \frac{h}{c^2}+\left(\frac{4}{3}-\frac{4}{9}h+\frac{1}{9}h^2\right)F_h(h)+\frac{h}{1-h}\log h \,,\nn\\
H_m(h)=&-\frac{h}{h-1}\log h +\frac{c^2 h}{h-c^2}\log \frac{h}{c^2}-\frac{s^2}{18 c^2}h+\left(\frac{h^2}{9}-\frac{4h}{9}+\frac{4}{3}\right)F_h(h)\nn\\
&\qquad-(c^2-s^2)\left(\frac{h^2}{9c^4}-\frac{4h}{9c^2}+\frac{4}{3}\right)F_h\left(\frac{h}{c^2}\right)\,.
\label{higgsdependence}
%H_A(h)=&\frac{h c^2}{h-c^2}\log \frac{h}{c^2}-\frac{8h}{9(h-1)}\log h+ \left({\textstyle\frac{4}{3}-\frac{2}{3}h+\frac{2}{9}h^2}\right)F_h(h)-\left({\textstyle\frac{4}{3}-\frac{4}{9}h+\frac{1}{9}h^2}\right)F_h'(h)-\frac{h}{18}\,,\nn\\
%H_R(h)=&-\frac{h}{18}+\frac{c^2}{1-\frac{c^2}{h}}\log \frac{h}{c^2}+\left(\textstyle{\frac{4}{3}-\frac{4}{9}h+\frac{1}{9}h^2}\right)F_h(h)+\frac{h}{1-h}\log h \,,\nn\\
%H_m(h)=&-\frac{h}{h-1}\log h +\frac{c^2 h}{h-c^2}\log \frac{h}{c^2}-\frac{s^2}{18 c^2}h+\left(\textstyle{
%\frac{4}{3}-\frac{4}{9}h+\frac{1}{9}h^2}\right)F_h(h)\nn\\
%&\qquad-(c^2-s^2)\left(\textstyle{\frac{1}{9}h^2/c^4-\frac{4h}{9c^2}+\frac{4}{3}}\right)F_h\left(\frac{h}{c^2}\right)\,.
\end{align}
These were obtained by dropping the Higgs mass independent terms 0.18, 0.03, 0.52 from the identically named functions in \cite{Novikov:1992rj}. The functions $F_h(h)$ and $F_h'(h)$ are given by\footnote{Note that $F_h'$ is not the derivative of $F_h$. Note also that the $F_h$ is given with a misprint in (F.7) of \cite{Novikov:1992rj}; the misprint is corrected in \cite{Novikov:1995vu}.}
\begin{align}
F_h(h)&=1+\left(\frac{h}{h-1}-\frac{1}{2}h\right)\log h+
\begin{cases}
 h\sqrt{1-\frac{4}{h}}\log{\left(\sqrt{\frac{h}{4}-1}+\sqrt{\frac{h}{4}}\right)}\,,&h>4\,,\\
- h\sqrt{\frac{4}{h}-1}\arctan{\sqrt{\frac{4}{h}-1}}\,,&h<4\,,
\end{cases}\nn\\
F_h'(h)&=-1+\frac{h-1}{2}\log h+
\begin{cases}
 (3-h)\sqrt{\frac{h}{h-4}}\log{\left(\sqrt{\frac{h}{4}-1}+\sqrt{\frac{h}{4}}\right)}\,,&h>4\,,\\
 (3-h) \sqrt{\frac{h}{4-h}}\arctan{\sqrt{\frac{4}{h}-1}}\,,&h<4\,.
\end{cases}
\end{align}
The experimental value $m_h=125$ GeV corresponds to $h\approx 1.89$, belonging to the $h<4$ branch of these formulas.

As a check, the Higgs mass dependence of \reef{eq:epsHiggs} agrees with the Higgs mass dependence of the full $\vareps$'s given in \cite{Novikov:1992rj}. On the other hand, \reef{eq:epsHiggs} also contains constants and logarithmically divergent terms of the Higgs boson contributions. These terms will be relevant in the composite Higgs discussion below. In particular the constant terms could not be easily extracted from \cite{Novikov:1992rj}, which gives only the total result for $\vareps$'s including the pure gauge boson loops, but not the separate Higgs contributions.

\section{$S$ parameter in the Minimal Composite Higgs Model}
\label{sec:MCHM}
\subsection{General idea}
\label{sec:genidea}
Let us formulate more precisely the problem that we want to solve. We are given a strong sector realizing an $SO(5)/SO(4)$ spontaneous symmetry breaking. The sector has a multiplet of global currents $J_\mu^A$. We assume that the two point functions of these currents are known at all energies (either measured or solved). This is before the strong sector is coupled to the SM gauge fields. Once these latter couplings are turned on, we expect that the backreaction on the current-current spectral densities at high energies will be negligible, while at low energies it will be to some extent universal. So we expect that a formula must exist which computes $\vareps_3$, at the lowest nontrivial order $O(g^2)$, in terms of the UV tail of the strong sector spectral densities, plus a universal IR piece. In this section we will present such a formula.

The computation proceeds in two steps. The first step is to replace the full strong sector by a low energy effective theory described by the Lagrangian
\beq
\frac 12(D_\mu \Phi)^2+O_S\,,\quad O_S=\frac{2 c(\mu^2)}{f^2} B_{\mu\nu} W^a_{\mu\nu} \Phi^t T^{aL} T^{3R} \Phi\,.
\label{eq:SILH}
\eeq
The coefficient $c(\mu^2)$ will have to be determined by a careful matching procedure; it will depend on the matching scale as indicated. In this part of the computation, the SM gauge fields $W_\mu$ and $B_\mu$ can be viewed as classical external sources introduced to measure the current correlation functions. The matching scale will be chosen in the range $m_h\ll\mu\ll m_\rho$, so that the Higgs boson mass can be neglected during the matching. That's why the Higgs mass term has been omitted from the Lagrangian \reef{eq:SILH}

The second step is to compute $\vareps_3$ starting from \reef{eq:SILH}. Now we have to make gauge bosons dynamical. We also have to take into account that the Higgs boson acquires a mass. The $\vareps_3$ gets a contribution from 
\begin{itemize}
\item the SM gauge boson and fermion loops, vertex and box corrections,
\item the Higgs boson loops, the second particle being a goldstone or a gauge boson,
\item the contact contribution from $O_S$.
\end{itemize}
The first group of contributions is identical in the SM and MCHM and will cancel upon computing the $S$ parameter.
The second contribution equals $\vareps_{3,\text{Higgs}}$ computed in section \ref{sec:SM}, up to a factor $a^2$ reflecting the reduction in the Higgs boson couplings.
As we will see, the regularization scale dependence of the Higgs contribution will cancel with that of $c(\mu^2)$, and the total $S$ parameter will be UV finite, and $\mu$-independent as it should.

\subsection{Strong sector current two point function}
\label{currents}
We start with a general discussion of the $SO(5)$ current-current correlation function (before coupling to the SM gauge bosons).
The form of this correlator is constrained by current conservation, $SO(5)$ symmetry, and the assumed symmetry breaking pattern $SO(5)/SO(4)$. The most general expression consistent with all the constraints is \cite{Agashe:2004rs}:
\beq
\langle J_\mu^A(q)J_\nu^B(-q)\rangle=-i P_{\mu\nu}\left[\delta^{AB}\,\Pi_0(q^2)+\widetilde\Phi^t T^A T^B \widetilde\Phi\,\Pi_1(q^2)+(T^A)_{ij} (T^B)_{kl}\eps^{ijklm}\widetilde\Phi^m\,\Pi_2(q^2)\right]\,.
\label{eq:JJ}
\eeq
% SR: 19.11.12 Pi_0 and Pi_1 in this formula are normalized exactly as in Agashe-Cotino-Pomarol, the factor 1/2 in their formula accounted for by the 1/2 appearing when Taylor-expanding <exp(i\int A. J)>
Here $\widetilde \Phi\equiv \Phi/f$ is the order parameter for the $SO(5)/SO(4)$ breaking, $P_{\mu\nu}=\eta_{\mu\nu} -{q_\mu q_\nu}/{q^2}$, and $T^A$ are the $SO(5)$ generators in the fundamental representation (see Appendix). The term proportional to the $\eps$-tensor, not mentioned in \cite{Agashe:2004rs}, is forbidden if the global symmetry is $O(5)$ rather than $SO(5)$. It breaks $P_{LR}$ parity and would give opposite sign contributions to the $SU(2)_L$ and $U(1)_Y$ gauge couplings once interactions with the SM gauge bosons are turned on; see Eq.~\reef{eq:Pi+-} below. 

The UV behavior of the $\Pi_i(q^2)$ form factors in \reef{eq:JJ} can be connected with the current-current OPE (see section 4.3 of \cite{Orgogozo:2011kq} for an analogous discussion in the $SU(2)\times SU(2)$ case). The OPE takes the form
\beq
J_\mu^A(x)J_\nu^B(0)=(\eta_{\mu\nu} \del^2-\del_\mu\del_\nu)\biggl[\frac{C_J\delta^{AB}}{(x^2)^{2}}+\sum_\calO \frac{ \Gamma^{AB}_C\calO^C(0)}{(x^2)^{2-\Delta_\calO/2}}+\ldots\biggr]\,,
\label{eq:OPE0}
\eeq
where we singled out scalars $\calO$ which get vevs when the $SO(5)$ symmetry breaks spontaneously, and the $\Gamma^{AB}_C$ are the Clebsch-Gordan coefficients. The `central charge' constant $C_J$ is a characteristic of the UV theory.\footnote{This constant is a measure of the number of degrees of freedom in the UV theory. If the global symmetry is weakly gauged, the CFT contribution to the one-loop beta-function will be proportional to $C_J$.} The $\ldots$ in the OPE contains fields of nonzero spin (including derivatives of scalars), which cannot get vevs if the Lorentz invariance is to be preserved. Notice that the OPE is written in the UV operator basis, where the EWSB sector is assumed to approach a scale-invariant fixed point.

Group theory restricts $\calO$ to transform in one of the representation appearing in the symmetric product of two adjoints of $SO(5)$: singlet $\bf 1$, traceless symmetric two-tensor $\bf 14$, fundamental $\bf 5$, and $\bf 35$ (which is $[2,2]$ in the Young tableau notation). The scalars in the first three channels will generically get vevs, whose orientation is determined by the order parameter:
\beq
\langle \calO\rangle\sim m_\rho^{\Delta_\calO}\times \begin{cases}1\,,&\calO\in\mathbf{1},\\
\widetilde\Phi^i\widetilde\Phi^j-\textstyle\frac{1}{5}\delta^{ij}\,,&\calO\in\mathbf{14},\\
\widetilde\Phi^i\,,&\calO\in\mathbf{5},\end{cases}
\eeq
while it can be shown that $\bf 35$ cannot get a vev consistent with a breaking to $SO(4)$.

The UV asymptotics of $\Pi_i(q^2)$ at $-q^2\gg m_\rho^2$ will be related to the lowest scalar dimension in the three vev channels:\footnote{For $\Pi_0$ actually the unit operator dominates, and so this equation describes the subleading asymptotics.}
\beq
\Pi_i(q^2)\sim q^2\left({-q^2}/{m_\rho^2}\right)^{-\Delta_{\calO_i}/2},\qquad \calO_0\in\mathbf{1},\calO_1\in\mathbf{14},\calO_2\in\mathbf{4}\,.
\eeq
That $\calO_2$ controls $\Pi_2$ is obvious because the Clebsch-Gordan coefficient is precisely the $\eps$-tensor appearing in \reef{eq:JJ}. 
That the vev of $\calO_1$ controls $\Pi_1$ is not immediately obvious but can be checked easily using the explicit Clebsch-Gordan coefficient and comparing with the prefactor in  \reef{eq:JJ}.

\subsection{Spectral densities}
\label{sec:dens}
Let us now consider separately the two point functions of the broken and unbroken currents (no summation in $a,\hat{a}$):
\begin{align}
\left\langle J^{\hat{a}}_{\mu}(q)J^{\hat{a}}_{\nu}(-q)\right\rangle &=- iP_{\mu\nu} \Pi_-(q^2)\,,\\
\left\langle J^{a\ L/R}_{\mu}(q)J^{a\ L/R}_{\nu}(-q)\right\rangle &=- iP_{\mu\nu} \Pi^{L/R}_+(q^2)\,.
\label{eq:defPipm}
\end{align}
Using the explicit expressions for the broken and unbroken generators given in the Appendix, the corresponding polarization operators can be expressed in terms of $\Pi_{0,1,2}$ \cite{Agashe:2004rs}. In particular, at the aligned vacuum ($\theta=0$) we get:
\beq
\Pi^{L/R}_+=\Pi_0\mp 2\Pi_2,\qquad\Pi_-=\Pi_0+\frac 12 \Pi_1\,\qquad (\theta=0)\,.
\label{eq:Pi+-}
\eeq

The point of introducing this new set of form factors is that they have positive spectral densities, being related to the two point functions of a current with itself. To write the corresponding dispersion relation, we define:
 {\begin{align}
  \Pi_-(q^2)&=-\frac 12 f^2+q^2 \widetilde \Pi_-(q^2)\,,\\
 \Pi^{L/R}_+(q^2)&=q^2 \widetilde \Pi^{L/R}_+(q^2)\,.
 \end{align}
 }
 Here we subtracted the goldstone pole in the broken current channel. The spectral densities
\beq 
\rho_{-}(s)=\frac{1}{\pi}\text{Im}\,\widetilde{\Pi}_{-}(s)
\eeq
and analogously defined $\rho^{L,R}_{+}$ are all positive. 
%tilded formfactors can now be represented by non-subtracted dispersion relations:
% \beq
%\widetilde{\Pi}_-(q^2)=\int ds \frac{\rho_-(s)}{s-q^2+i\eps},\ \ \ \ >0,
%\eeq
%and analogously for $\Pi^{L/R}_+$. The spectral densities $\rho_{-}$, $\rho^{L,R}_{+}$ are positive.

The $S$ parameter will be related to the left-right vacuum polarization:
\beq
\left\langle J^{3L}_{\mu}(q)J^{3R}_{\nu}(-q)\right\rangle =- iP_{\mu\nu}\Pi_{LR}(q^2)\,.
\label{eq:LR}
\eeq
Using the explicit generators and evaluating the RHS of \reef{eq:JJ} on the rotated vacuum \reef{eq:mis}, we find  \cite{Agashe:2004rs}
\beq
\Pi_{LR}(q^2)=\frac{v^2}{4}+q^2 \widetilde\Pi_{LR}(q^2),\qquad \widetilde\Pi_{LR}(q^2)=-\frac{v^2}{4f^2}\widetilde\Pi_{1}(q^2),
\eeq

From \reef{eq:Pi+-}, we can express $\Pi_{1}$ as:
\beq
\Pi_1=2 \Pi_- -\Pi^L_+  -\Pi^R_+\,.
\label{eq:Pi1}
\eeq
 Representing the formfactor $\Pi_1$ via non-subtracted dispersion relation, we obtain:
\beq
\widetilde{\Pi}_{LR}(q^2)=\frac{v^2}{4f^2}\int ds \frac{\rho(s)}{s-q^2+i\eps},\qquad \rho(s)=2[\rho_+(s)-\rho_-(s)]\,,\qquad \rho_+(s)=\frac 12[\rho^L_+(s)+\rho^R_+(s)]\,. 
\label{eq:piLR}
\eeq

Recall that in the Higgsless case of an $SU(2)_L\times SU(2)_R$ global symmetry spontaneously broken to $SU(2)_V$ there is an analogous formula:
\beq
\widetilde{\Pi}_{LR}(q^2)=\frac{1}{4}\int ds \frac{\rho_V(s)-\rho_A(s)}{s-q^2+i\eps}\,,
\eeq
where $\rho_{V/A}(s)>0$ are the spectral densities for the vector and axial current two point functions. In both cases the spectral density of unbroken (resp.~broken) generators enters with a positive (resp.~negative) sign. 

Notice that in the Higgsless case the parity $P_{LR}$ happens to coincide with the grading $R$ reversing the signs of the broken generators, while for $SO(5)/SO(4)$ these are two different automorphisms of the algebra. In the $SO(5)/SO(4)$ case the grading $R$ is a discrete symmetry of the strong sector, being an element of the unbroken $SO(4)$, while $P_{LR}$ belongs to $O(4)$ but not $SO(4)$ and may or may not be a true symmetry. It is however an accidental symmetry of the $SO(5)/SO(4)$ sigma-model at the two-derivative order (the first term in \reef{eq:leeff} in the limit of turned off gauge couplings). See \cite{Contino:2011np} for a lucid discussion of $R$ and $P_{LR}$.

We finish this section with comments about the IR and UV asymptotics of the spectral density $\rho(s)$, which will enter our final result. Just like in the Higgsless case, for $s\ll m_\rho^2$ the spectral density is dominated by the two-goldstone state and is nearly flat:
\beq
\rho(s)\approx 2\rho_+(s)\approx 1/(48\pi^2)\qquad(s\ll m_\rho^2)\,. 
\label{eq:flat}
\eeq
The actual value $1/(48\pi^2)$ is universal, since the coupling of two-goldstone state to the currents is fixed by the symmetry. The $\rho_-(s)$ part of the spectral density gets contribution from the three-goldstone states and, due to phase space reasons, is expected to be suppressed by $s/m_\rho^2$ in this limit. Eq.~\reef{eq:flat} will be crucial for the $\mu$-independence of our final $S$ parameter formula.

For $s\sim m_\rho$ the density will have complicated non generic structure reflecting the (even and odd grading) resonance spectrum. Finally, according to the discussion in the previous section, in the deep UV it will asymptote as 
\beq
\rho(s)\sim (s/m_\rho^2)^{-\Delta_{\calO_1}/2}\qquad (s\gg m_\rho^2)\,.
\label{eq:anom}
\eeq

The first and second Weinberg sum rules \cite{Weinberg:1967kj} in this context would take the form:
 \beq
 \int_0^\infty ds\,\rho(s)=f^2\,,\qquad \int_0^\infty ds\,s\,\rho(s)\stackrel{?}{=}0\,.
 \label{eq:WSR}
 \eeq
These equations follow by expanding $\Pi_1(q^2)$ at large $q^2$ and setting the $O(1)$ and $O(1/q^2)$ coefficients to zero. 
 Notice that this is legitimate only as long as $\Delta_{\calO_1}>2(4)$ for the first (second) sum rule. This remark is relevant for the models {\it \`a la} Conformal Technicolor \cite{Luty:2004ye}, which can also realize a pseudo-goldstone Higgs boson \cite{Galloway:2010bp}. A request of a scalar with the SM Higgs doublet quantum numbers and a small anomalous dimension is imposed in these models. This request necessitates the presence of low-dimension singlet scalars and even lower-dimension symmetric traceless tensors \cite{Rattazzi:2008pe,Rychkov:2009ij,Rattazzi:2010yc,Vichi:2011ux,Poland:2011ey}. The best current numerical bounds \cite{Vichi:2011ux,Poland:2011ey} are quite restrictive and imply that the second Weinberg sum rule is invalid in these scenarios.\footnote{The published bound in Fig.~6 of \cite{Poland:2011ey} concerns the case of an $SO(4)$ global symmetry, but an almost as strong bound holds for the $SO(5)$ case.} Incidentally, this also means that the Higgs boson mass in these scenarios is UV-sensitive, the contribution from the SM gauge boson loops (Eqs.~(19,22) in \cite{Agashe:2004rs}) being divergent.\footnote{We are grateful to Michele Redi for emphasizing this connection to us.}
 
 On the other hand, in the composite Higgs models where flavor is realized using the idea of partial compositeness (rather than {\it \`a la} Conformal Technicolor), there are no known \emph{a priori} constraints on the dimension of $\calO_1$. In those models both Weinberg sum rules might be satisfied. This is definitely the case in AdS realizations. See 
\cite{Marzocca:2012zn,Pomarol:2012qf} for recent applications of the Weinberg sum rules in the phenomenology of such models.

\subsection{Matching}

We now proceed with the first step in the computation of the $S$ parameter---matching the UV complete, strong sector theory to the effective theory described by the Lagrangian \reef{eq:SILH}. The matching will determine the coefficient $c(\mu^2)$ in $O_S$. To perform the matching, we have to pick an observable sensitive to this coefficient and compute it in the full and in the effective theory. The natural observable is the $J^{3L}J^{3R}$ current-current vacuum polarization, $\widetilde\Pi_{LR}(q^2)$ discussed in the previous section. We will evaluate it at $m_h^2\ll |q^2| \ll \mu^2$ (to be within the range of the effective theory) and at $q^2<0$ spacelike. In full theory $\widetilde\Pi_{LR}$ is given by the dispersion relation Eq.~\reef{eq:piLR}. We can simplify that equation by splitting the integral into two parts: above and below $\mu^2$. In the first part the density is very well approximated by the two-goldstone density \reef{eq:flat}. Up to $O(q^2/\mu^2)$ corrections we obtain:\footnote{Note added (April 2015): Notice that it would be wrong to think of $\mu$ in this computation as a regulating scale. This observable is finite and does not need a regulator in the full theory. We are just rewriting the exact expression for $\widetilde\Pi_{LR}(q^2)$ in the form explicitly exhibiting $\log{ \mu^2}/({-q^2})$ in preparation for the matching with the computation in the effective theory where $\mu$ will indeed play a role of the scale in dimensional regularization.}
\beq
\widetilde\Pi_{LR}(q^2)=\frac{v^2}{4f^2}\left[\frac{1}{48\pi^2}\log\frac{ \mu^2}{-q^2}+\int_{\mu^2}^\infty ds\,\frac{\rho(s)}{s}\right]\,.
\label{eq:full}
\eeq

We now have to compute the same observable in the effective theory. 
The current-current correlation function will be the difference of the goldstone-goldstone and Higgs-goldstone loops evaluated at the external momentum $q$. We also have to include the contact contribution from $O_S$. Since we are assuming $-q^2\gg m_h^2$, the Higgs boson mass can be neglected. We do this computation in dimensional regularization at the scale $\mu$. Up to $O(q^2/\mu^2)$ corrections, we get\footnote{The finite term appears from an integral over the Feynman parameter: $6\int_0^1 dx\,(x-x^2)\log (x-x^2)=5/3$.}
{\begin{align}
\widetilde\Pi_{LR}(q^2)&=\frac{v^2}{f^2}c(\mu^2)+\frac{1}{192\pi^2}\Bigl(N_\eps+\log\frac{ \mu^2}{-q^2}+\frac{5}{3}\Bigr)_{\text{goldstones}}-\frac{a^2}{192\pi^2}\Bigl(N_{\eps}+\log\frac{\mu^2}{-q^2}+\frac{5}{3}\Bigr)_{\text{Higgs}}\nn\\
&=\frac{v^2}{f^2}\left[c(\mu^2)+\frac{1}{192\pi^2}\left(N_\eps+\log\frac{\mu^2}{-q^2}+\frac{5}{3}\right)\right]\,.
\end{align}
}
Requiring that the result agree with the full theory answer \reef{eq:full} fixes $c(\mu^2)$:
\beq
c(\mu^2)=\frac{1}{4}\int_{\mu^2}^\infty ds\,\frac{\rho(s)}{s}-\frac{1}{192\pi^2}\left(N_{\eps}+\frac{5}{3}\right)
\eeq
Notice that the dependence on $q^2$ cancels as it should, since we can do the matching at any $q^2\ll\mu^2$ and the result must be the same.

\subsection{Computing $\vareps_3$ for Composite Higgs}

Now that the matching is done, we can use the effective theory to compute the low-energy observable $\vareps_3$. We couple the effective theory to dynamical gauge bosons, and also add the Higgs boson mass. The backreaction of these changes on the effective theory parameters measured at the matching scale $\mu$ is negligible if $m_h,m_W\ll \mu$ as we assume. 

As explained at the end of section \ref{sec:genidea}, all the contributions to the $\vareps_3$ having to do with the loops of the $W$ and $Z$ (this includes vertex and box corrections) will be the same as in the SM. The Higgs loop contribution will be reduced by $a^2$, and there will be a new contribution proportional to $c(\mu^2)$. In detail, we have:
{\begin{align}
\vareps_{3}\text{(SM)}&=\vareps_{3,W/Z}+\vareps_{3,\text{Higgs}}\,,\\
 \vareps_3\text{(MCHM)}&=\vareps_{3,W/Z}+a^2\vareps_{3,\text{Higgs}}+g^2\frac{v^2}{f^2}c(\mu^2)\,.
\end{align}
}
Here $\vareps_{3,\text{Higgs}}$ is the SM Higgs contribution which we computed in Eq.~\reef{eq:epsHiggs}, and $c(\mu^2)$ has been computed via matching in the previous section. The $\vareps_{3,W/Z}$ is the only term which we have not computed but it cancels when we form the difference, which is the $S$ parameter in our normalization:
\beq
\hat{S}=\vareps_3\text{(MCHM)}-\vareps_{3}\text{(SM)}=\frac{v^2}{f^2}\left[-\vareps_{3,\text{Higgs}}+g^2 c(\mu^2)\right]\,.
\eeq
Notice that we define the $S$ parameter \textbf{with respect to the SM with the same reference value of the Higgs mass} as that in the MCHM.

Substituting the values of $\vareps_{3,\text{Higgs}}$ and $c(\mu^2)$, the $N_\eps$ dependence cancels out, and we get a UV-finite answer:
\beq
\boxed{\ \hat{S}=g^2\frac{v^2}{f^2}\left[-\frac{3}{64\pi^2}\left[H_A(h)-H_R(h)\right]+\frac{1}{192\pi^2}(\log\frac{\mu^2}{m_Z^2}-1)+\frac{1}{4}\int_{\mu^2}^\infty ds\,\frac{\rho(s)}{s}
\right]\ }
\label{eq:main}
\eeq
This formula is the main result of this paper. It computes the $S$ parameter of the MCHM at the leading order in $g^2$. We remind that the scale $\mu$ used to separate the IR and the UV parts of the final answer should be chosen in the interval
\beq
m_h\ll\mu \ll m_\rho\,.
\eeq
The total then does not depend on $\mu$ as the spectral density in this interval is nearly flat; see Eq.~\reef{eq:flat}.

The finite separation of scales $m_h/m_\rho$ causes an imprecision in the matching process and in our final result, which can be estimated as follows. The (relative) error in the first matching step, which neglects both $m_h$ and $m_W$ (keeping gauge fields non-dynamical), can be estimated as (we pick the larger mass of $m_h,m_W$)
\beq
err_1=O(m_h^2/\mu^2)\,.
\eeq
To minimize this error the matching should be done at as high a scale $\mu$ as possible. However, in the second step, when we combine the low energy contribution to $\hat S$ with the matching coefficient $c(\mu^2)$, the opposite request is necessary. This is because the $\mu$ dependence cancels modulo $O(s/m_\rho^2)$ deviations of the spectral density from its universal 
low energy form $1/(48\pi^2)$. So the error in the second step can be estimated as
\beq
err_2=O(\mu^2/m_\rho^2)\,.
\eeq
The sum $err_1+err_2$ is minimized for $\mu\sim(m_\rho m_h)^{1/2}$ for which it is
\beq
err_{\rm tot}=O(m_h/m_\rho)\,.
\eeq
We believe that this is a fair estimate of the relative accuracy of our formula \reef{eq:main}\,.

\section{Examples and benchmarks}
 \label{sec:toy}
 
To take full advantage of the accuracy that Eq.~\reef{eq:main} provides, one needs to know the non-generic part of the spectral density $\rho(s)$, i.e.\ across the resonance region and into the UV. If the MCHM scenario is realized in nature, this knowledge may come from future experiments and theoretical computations based on a high energy description of the EWSB sector. Here we will limit ourselves to discuss toy model examples. In particular we would like to see how Eq.~\reef{eq:main} compares to the simpler treatment of Ref.~\cite{Barbieri:2007bh} discussed in the introduction. 
%We therefore consider various toy model spectral densities. 

\subsection{Heavy vs light Higgs}

The main purpose of our first example is to confront the heavy and light Higgs limits of our result. We therefore pick an extremely naive parametrization for the UV part of the spectral density. The density will be assumed flat at the level corresponding to Eq.~\reef{eq:flat} up to the mass of the first and only $R$-even resonance, where it has a $\delta$-function singularity and drops to zero:
\beq
 \rho(s)=\frac{1}{48\pi^2}\theta(m_\rho^2-s)+ F_\rho^2\, \delta(s-m_\rho^2)\,.
\label{eq:toy1}
\eeq
We are implicitly assuming here the $P_{LR}$ symmetry, otherwise the resonances in $\rho^L_+$ and $\rho^R_+$ parts of the spectral density would not have the same mass. As discussed in Section \ref{sec:dens}, in any theory the low energy limit of $\rho(s)$ is given by $1/(48\pi^2)+O(s/m_\rho^2)$. Here we neglect the deviations from $1/(48\pi^2)$ for simplicity. Also, this being but a toy model, we do not discuss constraining its parameters by imposing the Weinberg sum rules (there is not even enough freedom to impose them both).

For the density \reef{eq:toy1} we would get the $S$ parameter:
\beq 
\hat{S}=g^2\frac{v^2}{f^2}\left[-\frac{3}{64\pi^2}\left[H_A(h)-H_R(h)\right]+\frac{1}{192\pi^2}(\log\frac{m_\rho^2}{m_Z^2}-1)+\frac{F_\rho^2}{4m_\rho^2}\right]
\eeq
The last term in this formula is what we called $\hat{S}_\text{UV}$ in Eq.~\reef{eq:SUV}. The estimate match if $F_\rho\sim f$, which is natural as we discuss below. To compare with Ref.~\cite{Barbieri:2007bh}, let us expand in the heavy Higgs limit: 
\beq
\hat{S}\approx g^2\frac{v^2}{f^2}\left[\frac{1}{96\pi^2}\Bigl(\log\frac{m_\rho}{m_h}-\frac{11}{12}\Bigr)+\frac{F_\rho^2}{4m_\rho^2}\right]\qquad
(m_h\gg m_Z)
\,.
\label{eq:HH}
\eeq
The logarithm in the first term is what we would get replacing $m_h\to m_\text{eff}$ with $\Lambda=m_\rho$. We also retained the finite term, which is the same $11/12$ one gets when subtracting the reference Higgs contribution in the Peskin-Takeuchi formula (see e.g.~Eq.~(4.10) of \cite{Orgogozo:2011kq}). In fact this limit of our formula could be derived starting from the Peskin-Takeuchi formula.

However, the observed value of the Higgs mass $m_h=125$ GeV lies beyond the range of validity of the heavy Higgs approximation. Evaluating the $H_{A,R}$ functions exactly, the $S$ parameter can be conveniently written down in the form:\footnote{Ref.~\cite{Marzocca:2012zn}, footnote 24, also discussed numerical impact of going beyond the heavy Higgs approximation.}
\beq
\hat{S}= g^2\frac{v^2}{f^2}\left[\frac{1}{96\pi^2}\Bigl(\log\frac{m_\rho}{125\text{ GeV}}-0.29\Bigr)+\frac{F_\rho^2}{4m_\rho^2}\right]\qquad
(m_h= 125\text{ GeV})
\,.
\label{eq:125}
\eeq
The difference between this equation and the heavy Higgs limit allows to appreciate the importance of finite terms that we computed. Notice that the prescription of \cite{Barbieri:2007bh} would correspond to dropping $-11/12$ in \reef{eq:HH} and $-0.29$ in \reef{eq:125}. We see that numerically the old prescription works even better for a light Higgs than for a heavy one.

\subsection{Vector Meson Dominance} 
Our second example is analogous to Vector Meson Dominance (VMD) as applied in QCD and in Higgsless models. It will contain one $R$-even and one $R$-odd resonance, playing a role similar to the $\rho$ and $a_1$ mesons. For the time being, we model the spectral density by:\footnote{One can improve spectral density modeling by including other two-particle intermediate states, e.g.~the goldstone-vector meson states as done in \cite{Pich:2012jv} for the Higgsless models. We will not attempt to do this here.}
\beq
\rho(s)=\frac{1}{48\pi^2}\theta(m_+^2-s)+F_+^2\delta(s-m_+^2)-F_-^2\delta(s-m_-^2)
\label{eq:rhoVMD}
\eeq
Again we are assuming $P_{LR}$, although this is not essential. Later on, we will see that this model also allows to replace the hard cutoff of the two-goldstone spectral density by a gentler mechanism. 
Analogously to the previous example, the above spectral density leads to:
\beq \hat{S}=g^2\frac{v^2}{f^2}\frac{1}{96\pi^2}\Bigl(\log\frac{m_+}{125\text{ GeV}}-0.29\Bigr)+\hat{S}_{\text{UV}}\qquad
(m_h= 125\text{ GeV})\,,
\eeq
where
\begin{align}
\hat{S}_{\text{UV}}&=\frac{g^2}{4}\frac{v^2}{f^2}\left[\frac{F_+^2}{m_+^2}-\frac{F_-^2}{m_-^2}\right]\ \stackrel{\text{WSR}}{\longrightarrow}\  {m_W^2}(m_+^{-2}+m_-^{-2})\,.
\label{eq:ifWSR}
\end{align}
The shown replacement happens if we assume that the two resonances saturate the two Weinberg sum rules \reef{eq:WSR}, i.e.~if\footnote{We neglect the subleading contribution of the two-Goldstone part of the spectral density to the sum rules.}
\beq
F_+^2-F_-^2=f^2, \qquad F_+^2 m_+^2-F_-^2 m_-^2=0\,.
\label{eq:WSR2}
\eeq
We see that in this case $\hat{S}_{\text{UV}}$ is positive and similar to the estimate in Eq.\reef{eq:SUV}. But, as discussed in section \ref{sec:dens}, the second Weinberg sum rule will not be satisfied in the models realizing flavor \emph{\`a la} Conformal Technicolor, in which case the UV contribution could then significantly deviate from the RHS of Eq.~\reef{eq:ifWSR}.

To explain how the VMD spectral density \reef{eq:rhoVMD} arises, one can build a Lagrangian for the resonances. We use the CCWZ formalism \cite{Coleman:1969sm,Callan:1969sn}; our notation follows closely \cite{Contino:2011np}. We parametrize the goldstones of $SO(5)/SO(4)$ as $U(\pi)=e^{i \sqrt{2}\pi(x)/f}$, where $\pi(x)=\pi^{\hat{a}}(x)T^{\hat{a}}$.  
Under an $SO(5)$ transformation $g$:
\beq
U(\pi)\rightarrow g U(\pi) h^{\dagger}(\pi,g), \ \ h\in SO(4)\,.
\eeq
The projections onto the broken/unbroken generators of the Cartan-Maurer one-form
\beq
-iU^{\dagger}(\pi)\partial_\mu U(\pi) = d^{\hat{a}}_\mu T^{\hat{a}}+E^a_\mu T^a\equiv d_\mu+E_\mu
\eeq
transform under $SO(5)$ transformations as follows:
\beq
\begin{split}
d_\mu &\rightarrow h(\pi,g) d_\mu h^{\dagger}(\pi,g)\,,\\
E_\mu &\rightarrow h(\pi,g) E_\mu h^{\dagger}(\pi,g)-i h(\pi,g) \partial_\mu h^{\dagger}\,.
\end{split}
\eeq
We see that the $d_\mu$ transforms homogeneously, and so will be a field strength constructed out of $E_\mu$:
\beq
E_{\mu\nu}=\partial_{\mu}E_\nu-\partial_\nu E_\mu +i[E_\mu,E_\nu]\,.
\eeq
To keep the notation simple, we gauge the whole $SO(5)$ and then freeze the unwanted degrees of freedom. We thus define $A_\mu=A^{\hat{a}}T^{\hat{a}}+A_\mu^a T^a$ which transforms as a gauge field under local $SO(5)$ transformations, and denote by $F_{\mu\nu}$ the corresponding field strength. 
We then define the structure:
\beq
f_{\mu\nu}=U^{\dagger}F_{\mu\nu}U=(f_{\mu\nu}^-)^{\hat{a}}T^{\hat{a}}+(f_{\mu\nu}^+)T^a\equiv f_{\mu\nu}^- + f_{\mu\nu}^-\,,
\eeq
where both $f_{\mu\nu}^-$ and $f_{\mu\nu}^-$ transform homogeneously.

To introduce the two resonances, we use the antisymmetric tensor formalism \cite{Ecker:1988te}, and make them transform linearly under the unbroken subgroup.
We take an $R$-even resonance multiplet $\rho_+^{\mu\nu}\equiv\rho_{+,a}^{\mu\nu}T^a$ in the adjoint of $SO(4)$ and an $R$-odd $\rho_-^{\mu\nu}\equiv\rho_{-,\hat{a}}^{\mu\nu}T^{\hat{a}}$ in the fundamental. Their transformations under $SO(5)$ can be written as:
\beq
\rho_\pm^{\mu\nu}\to h(\pi,g)\rho_\pm^{\mu\nu}h^\dagger(\pi,g)\,.
\eeq
The kinetic/mass terms take the form:
\beq
\mathcal{L}_{kin}=\frac{1}{2}\left\langle\nabla_\mu \rho_{\pm}^{\mu\nu} \nabla_\sigma \rho_{\pm}^{\sigma\nu}\right\rangle+\frac{1}{4}m_\pm^2 \left\langle \rho_\pm^{\mu\nu}\rho_\pm^{\mu\nu} \right\rangle \,,
\label{eq:Lkin}
\eeq
where $\nabla_\mu \rho\equiv \partial_\mu \rho +i [E_\mu,\rho]$.

The $O(p^2)$  interaction lagrangian relevant to study the $S$ parameter is:
\beq
\mathcal{L}_{int}=\frac{G_+}{2}\left\langle \rho_{+}^{\mu\nu}[d_\mu,d_\nu]\right\rangle+\frac{F_+}{2}\left\langle \rho_{+}^{\mu\nu}f_{+}^{\mu\nu}\right\rangle+\frac{F_-}{2}\left\langle \rho_{-}^{\mu\nu}f_{-}^{\mu\nu}\right\rangle\,.
\label{eq:Lint}
\eeq
If we neglect $G_+$, this lagrangian reproduces the spectral density \reef{eq:rhoVMD}, except that the two-goldstone part is not cut off at $m_+$. 

The subsequent discussion follows closely section 4.2 of \cite{Orgogozo:2011kq}. To see how the spectral density gets cut off, let's turn on $G_+$. The $R$-even resonance gives a tree level contribution to the two-goldstone form factor:
\beq
\langle \pi^{\hat{a}}(p')\left.\right|J^\mu_c\left|\right. \pi^{\hat{b}}(p)\rangle=C^{\hat{a}\hat{b}c}\mathcal{F}(q^2)(p+p')^\mu\qquad(q=p'-p)\,,
\eeq
where
\beq
\mathcal{F}(q^2)=1-\frac{F_+ G_+}{f^2}\frac{q^2}{q^2-m_+^2}\,.
\eeq
The first term corresponds to the contact contribution from the sigma-model, while the second one comes from the tree level resonance exchange.
The goldstones being composite, this form factor should go to $0$ as $q^2\rightarrow\infty$.
The premise of VMD is to achieve this with one $R$-even resonance, which requires
\beq
F_+ G_+ = f^2\,.
\label{eq:ffrel}
\eeq

Furthermore, the form factor $\mathcal{F}(q^2)$ affects the two-goldstone state contribution to the spectral density $\rho_+(s)$:
\beq
\rho_+^{\pi\pi}=\frac{1}{96\pi^2}\rightarrow \frac{1}{96\pi^2}\left|\mathcal{F}(s)\right|^2\,.
\eeq
Introducing the finite decay width to two goldstones in the resonance propagator
\beq
\Gamma_+=\frac{G_+^2 m_+^3}{48\pi f^4}
\eeq
and using \reef{eq:ffrel}, we get the modified spectral density:
\beq
\rho_+(s)=\frac{1}{96\pi^2}\frac{m_+^4}{(s-m_+^2)^2+\Gamma^2_+ m_+^2}\,.
\eeq
We now see how the flat spectral density is cut off:
at energies above the $R$-even resonance mass, the interference between the two contributions to the two-goldstone form factor is destructive and for $s \gg m_+^2$, $\rho_+(s)$ drops fast as $1/s^2$. Of course this simple model is unable to reproduce the anomalous dimension fall off \reef{eq:anom}. 

Including the $R$-odd resonance contribution, the full spectral density is given by
\beq
\rho(s)=\frac{1}{48\pi^2}\frac{m_+^4}{(s-m_+^2)^2+\Gamma^2_+ m_+^2}-F_-^2\delta(s-m_-^2)\,.
\eeq
Eq.~\reef{eq:rhoVMD} can be seen as an approximation to this, valid in the limit $\Gamma_+/m_+\to0$.

Note that one could get an idea about the size of the coupling $G_+$ by studying unitarization of goldstone-goldstone scattering via the $\rho_+$-exchange (similarly to how it was done in Higgsless models in \cite{Barbieri:2008cc}, \cite{Orgogozo:2011kq}). The allowed range of $G_+$ would depend on $\xi\equiv v^2/f^2$, which determines the Higgs boson contribution to the scattering amplitude, $m_+$ and the energy scale up to which unitarization is desired. One could then determine $F_+$ from \reef{eq:ffrel} and get an idea about $S_{\text{UV}}$. The $F_-$ can be determined from the first Weinberg sum rule, and the mass of $\rho_-$ from the second one. Alternatively one may decide to relax the second sum rule and keep $m_-$ as a free parameter. We leave this numerical analysis to the future.

\subsection{Remark on the gauge formalism}

We finish this section with a brief comment concerning the $S$ parameter discussion of Ref.~\cite{Contino:2011np}. There spin one resonances with quantum numbers (\textbf{3},\textbf{1}) or (\textbf{1},\textbf{3}) under the unbroken $SU(2)_L\times SU(2)_R= SO(4)$ were introduced as gauge fields transforming non-homogeneously:
\beq
\varrho_\mu \rightarrow  h(\pi,g)\varrho_\mu h^\dagger(\pi,g)-ih(\pi,g)\partial_\mu h^{\dagger}(\pi,g)\,. 
\eeq  
Taking $\varrho=(\textbf{3},\textbf{1})$ for concreteness, they compute the $S$ parameter at tree level from the following Lagrangian:
\beq
\mathcal{L}_{\text{gauge}}=-\frac{1}{4g_{\varrho}^2}\varrho_{\mu\nu}^{a_L}\varrho^{a_L \mu\nu}
+\frac{m_{\varrho}^2}{2 g_{\varrho}^2}(\varrho^{a_L}_\mu-E_\mu^{a_L})^2
+\alpha_2 \left\langle\varrho^{\mu\nu} f_{\mu\nu}^+\right\rangle\,,
\label{eq:contino}
\eeq
where $\varrho_{\mu\nu}$ is the field strength constructed out of $\varrho_\mu$.
This gives them:
\beq
\hat{S}=\frac{m_W^2}{g_{\varrho}^2f^2}-4\alpha_2 \frac{m_W^2}{f^2}\,.
\eeq
This formula predicts that $S$ may become negative for sufficiently large positive $\alpha_2$. Ref.~\cite{Contino:2011np} did not use it in this regime, but only for moderate $\alpha_2>0$ where the $S$ parameter is reduced but is still positive. Still, from our point of view this formula is puzzling, because the dispersion relation predicts that the $R$-even resonances should under all circumstances give a non-negative contribution to $S$.

Ref.~\cite{Contino:2011np} noticed that parameter $\alpha_2$ in \reef{eq:contino} should satisfy the constraint 
\beq
\alpha_2<1/(g_\varrho g)\,,
\label{eq:maxalpha}
\eeq 
since otherwise when coupling to the SM gauge field and diagonalizing, one of the 
two kinetic terms comes out with a negative sign, signaling a ghost instability. This range still allows $S$ to become negative. There is also a logical problem with Eq.~(\ref{eq:maxalpha}). Since $\alpha_2$ is a parameter characterizing the strong sector, whatever consistency condition on it must be expressible in terms of other strong sector parameters and cannot depend on the SM gauge coupling $g$. The strong sector does not know to which theory we may decide to couple it; it is either consistent by itself or inconsistent.

In our opinion the resolution of the puzzle may lie elsewhere.\footnote{See \cite{Ecker:1989yg} for similar reasonings.} It comes from examining the full $q^2$ dependence of the current-current vacuum polarization. The gauge Lagrangian \reef{eq:contino} gives:
\beq
\Pi_+^L(q^2)=-\frac{q^2}{q^2-m_{\varrho}^2}\left[\frac{m_{\varrho}^2}{g_{\varrho}^2}-4\alpha_2 m_{\varrho}^2 +4\alpha_2^2 g_{\varrho}^2 q^2\right]\,.
\label{eq:Pi+Lg}
\eeq
and 
\beq
\Pi_-(q^2)=\Pi_+^R(q^2)=0\,.
\label{eq:Pi-+Rg}
\eeq
This leads to inconsistencies with the OPE. First, from the definitions of the form factors in Eq.~\reef{eq:Pi+-} all three of them should have the same leading large $q^2$ behavior, determined by $\Pi_0(q^2)$:
\beq
\Pi_+^{L/R}(q^2)\sim\Pi_-(q^2)\sim\Pi_0(q^2)\sim \gamma q^2\qquad (q^2\to\infty)\qquad\text{(must be)}\,,
\eeq
where the constant $\gamma$ is proportional to the central charge $C_J$ in \reef{eq:OPE0}. This is not what (\ref{eq:Pi+Lg},\ref{eq:Pi-+Rg}) imply. Another manifestation of the same problem is that if we use (\ref{eq:Pi+Lg},\ref{eq:Pi-+Rg}) to compute $\Pi_1$ via Eq.~\reef{eq:Pi1}, we find
\beq
\Pi_1(q^2)\sim 4\alpha_2^2 g_{\varrho}^2 q^2 \qquad (q^2\to\infty),
\eeq
in contradiction with the OPE since the unit operator does not contribute in this channel.

A way to fix these issues is to add the following contact terms to the gauge Lagrangian:
\beq
\Delta \mathcal{L}_{\text{gauge}}=\frac{1}{2}\alpha_2^2 g_\rho^2(\left\langle f_-^{\mu\nu} f_-^{\mu\nu} \right\rangle\ + \left\langle f_{+,R}^{\mu\nu} f_{+,R}^{ \mu\nu} \right\rangle\ -\left\langle f_{+,L}^{\mu\nu} f_{+,L}^{\mu\nu} \right\rangle)
\eeq  
The form factors computed with the corrected Lagrangian are:
\beq
\begin{split}
\Pi_+^L(q^2)&=-\frac{q^2}{q^2-m_{\varrho}^2}\left[\frac{m_{\varrho}^2}{g_{\varrho}^2}-4\alpha_2 m_{\varrho}^2 +4\alpha_2^2 g_{\varrho}^2 q^2\right]+2\alpha_2 g_\rho^2 q^2\,,\\
\Pi_+^R(q^2)&=\Pi_-(q^2)=-2\alpha_2^2 g_\rho^2 q^2\,,
\end{split}
\eeq
and
\beq
\Pi_1(q^2)=m_{\varrho}^2 \frac{q^2}{q^2-m_{\varrho}^2}\left(\frac{1}{g_{\varrho}}-2\alpha_2 g_{\varrho}\right)^2\,.
\eeq
They are all consistent with the OPE. The predicted $S$ parameter becomes non-negative as expected
\beq
\hat{S}=\frac{m_W^2}{f^2}\left(\frac{1}{g_{\varrho}}-2\alpha_2 g_{\varrho}\right)^2\ge 0\,.
\eeq

With our VMD lagrangian, $\Pi_1$ was directly well behaved and $S$ was positive. However, if one looks at the form factors $\Pi_+^{L/R}$ and $\Pi_-$ individually, one may notice that their UV behavior behavior is too soft, corresponding to the common value $\gamma=0$:
\beq
\Pi_+^{L}(q^2)=
\Pi_+^{R}(q^2)=-\frac{1}{4}\frac{q^2 F_+^2}{q^2-m_+^2}\,,\qquad 
\Pi_-^{}(q^2)=-\frac{1}{4}\frac{q^2 F_-^2}{q^2-m_-^2}\qquad \text{(VMD)}\,.
\eeq
To allow for $\gamma\ne 0$, we can add contact terms, which are now LR symmetric: 
\beq
\Delta \mathcal{L}_{\text{VMD}}=-\frac{\gamma}{4}(\left\langle f_-^{\mu\nu} f_-^{\mu\nu} \right\rangle\ +  \left\langle f_+^{\mu\nu} f_+^{ \mu\nu} \right\rangle)
\eeq
This gives every form factor the same $\gamma q^2$ behavior at infinity, restoring consistency with the $\Pi_0$ OPE. Unlike for the gauge Lagrangian, the $S$ parameter computation is not affected by these contact terms.

\section{Discussion and outlook}

In this paper we have shown how the $S$ parameter can be computed precisely in a theory where the light Higgs boson arises as a pseudo-goldstone boson belonging to a strong sector realizing a spontaneously broken global symmetry. Our main result is Eq.~\reef{eq:main}, which combines an IR contribution from the light Higgs boson with an UV contribution from spin one resonances. Our formula encodes the same physics as earlier treatments, but is much more accurate due to a careful matching procedure. Of course, to realize the full accuracy of the formula, the strong sector spectral density must be known. This knowledge may come from future experiments, Monte-Carlo simulations, or theoretical computations if and when a microscopic description of the strong sector becomes known. 

For simplicity, we considered here the minimal symmetry breaking pattern giving rise to a composite pseudo-goldstone Higgs boson: $SO(5)/SO(4)$.
However, our method is completely general and would work for more complicated breakings having extra goldstones, such as $SO(6)/SO(5)$ \cite{Gripaios:2009pe}, $SO(6)/SO(4)\times SO(2)$ \cite{Mrazek:2011iu} or $SO(7)/G_2$ \cite{Chala:2012af}. It would also work if the Higgs boson is not a goldstone but belongs to a strongly interacting sector so that its couplings are suppressed for reasons other than the sigma-model structure, such as the Randall-Sundrum model case considered in Ref.~\cite{Burdman:2008gm}.\footnote{We are leaving here aside the issue that the latter models don't explain the lightness of the Higgs boson.}

It would be very interesting to find a dispersion relation similar to the ones considered in this work but for the $T$ parameter. The difficulties in finding such a dispersion relation were already emphasized by Peskin and Takeuchi \cite{Peskin:1991sw}. At present it is not even known which strong sector observable would control its UV part. Currently the $T$ parameter in composite Higgs models is studied using the simplified models of vector-like fermionic resonances \cite{Barbieri:2007bh,Lodone:2008yy,Gillioz:2008hs,
Anastasiou:2009rv,Barbieri:2012tu}. Potentially large $O(g'^2)$ contributions were identified in Higgsless models by Feynman diagram calculations starting from VMD Lagrangians \cite{Barbieri:2008cc,Orgogozo:2011kq}. A dispersion relation could help understanding the theoretical uncertainties in these calculations.

\section*{Acknowledgements}

We are grateful to Adam Falkowski, Roberto Contino, Marco Serone and Michele Redi for useful comments and discussions. We are especially grateful to Roberto Contino for checking many of our computations and pointing out misprints in Section 4.3.
This work was supported in part by the European Program ``Unification in the LHC Era'', contract PITN-GA-2009-237920 (UNILHC), and by the \'Emergence-UPMC-2011 research program.

\appendix
\section{$SO(5)$ generators}\label{generators}
We use the $SO(5)$ generators in the fundamental representation normalized as $\langle T^A T^B\rangle \equiv \text{Tr}[T^A T^B]=1$. For the broken and the unbroken generators we use the explicit expressions \cite{Contino:2011np}:
\begin{align}
T^{\hat{a}}_{ij}&=\frac{i}{\sqrt{2}}\left(\delta^{\hat{a}i}\delta^{5 j}-\delta^{\hat{a}j}\delta^{5 i}\right),\\
T^{a\ L/R}_{ij}&=-\frac{i}{2}\left[\frac{1}{2}\epsilon^{abc}\left(\delta^{bi}\delta^{cj}-\delta^{bj}\delta^{ci}\right)\pm \left(\delta^{ai}\delta^{4j}-\delta^{aj}\delta^{4i}\right)\right]\,,
\end{align}
where $\hat{a}=1,...,4$, $a=1,2,3$ and $i,j=1,...,5$.

The structure constants are denoted by $[T^A,T^B]=i C^{ABC}T^C$. For the computation of the width $\Gamma_+$ and of the two-goldstone form factor, two useful relations are:
\beq
C^{\hat{a}\hat{b}e}C^{\hat{c}\hat{d}e}=-\frac{1}{2}\left(\delta^{ab}\delta^{cb}-\delta^{ac}\delta^{bd}\right)
\eeq 
and
\beq
\bigl\langle T^a \{T^{\hat{a}},T^{\hat{b}}\}\bigr\rangle=0\,.
\eeq

\bibliography{Biblio-Scomp2}{}
\bibliographystyle{utphys}

\end{document}